\documentclass[seceq]{ptptex}
\title{%        %You can use \\ for explicit line-break.
Particle Physics at the LHC Start%
}

\author{%       %Use \scshape for the family name.
Guido \textsc{Altarelli}%
}

\inst{%     %Affiliation, neglected when [addenda] or [errata].
Dipartimento di Fisica `E.~Amaldi', Universit\`a di Roma Tre\\
and INFN, Sezione di Roma Tre, I-00146 Rome, Italy\\ and \\ CERN, Department of Physics, Theory Unit, 
\\CH--1211 Geneva 23, Switzerland
}

\abst{%         %This abstract is neglected when [addenda] or [errata].
I present a concise review of the major issues and challenges in particle physics at the start of the LHC era. After a brief overview of the Standard Model and of QCD, I will focus on the electroweak symmetry breaking problem which plays a central role in particle physics today. 
The Higgs sector of the minimal Standard Model is so far just a mere conjecture that needs to be verified or discarded by the LHC. Probably the reality is more complicated. I will summarize the motivation for new physics that should accompany or even replace the Higgs discovery and a number of its possible forms that could be revealed by the LHC.
\\
\\
RM3-TH/10-10~~~~~~~~CERN-PH-TH/2010-249
}

\begin{document}

\maketitle

\section{Introduction}

In a few words the general map of particle physics at the LHC start is as follows. The Standard Model (SM) \cite{sprin} is a low energy effective theory (nobody can believe it is the ultimate theory). It happens to be renormalizable, hence highly predictive and is extremely well supported by the data. However, one expects corrections from higher energies, in particular already from the TeV scale (LHC!), and also from the GUT/Planck scales and possibly from some additional intermediate scales. But even as a low energy effective theory the SM is not satisfactory. In fact while QCD and the gauge part of the EW theory are well established, the Higgs sector is so far just a conjecture. Not only it needs an experimental verification but it introduces serious theoretical problems, like the hierarchy problem, that demand some form of new physics at the electroweak scale. 

The really good news is that finally the physics run at $3.5$ TeV per beam of the LHC is successfully going on. The current plan is to collect 1 $fb^{-1}$ at 7 TeV total c.o.m. energy by the end of 2011, and then, after a one year long shut-down, increase the energy up to the design figure of 14 TeV. The particle physics community eagerly waits for the answers from the LHC  to a number of big questions. As hinted above the main physics issues at the LHC, addressed by the ATLAS and CMS collaborations, will be: 1) the experimental clarification of the Higgs sector of the electroweak theory, 2) the search for new physics at the weak scale that, on conceptual grounds, one predicts should be in the LHC discovery range, and, hopefully, 3) the identification of the particle(s) that make the Dark Matter in the Universe, in particular if those are WIMPs (Weakly Interacting Massive Particles). In addition the LHCb detector is dedicated to the study of precision B physics, with the aim of going deeper into the physics of the Cabibbo-Kobayashi-Maskawa (CKM) matrix and of CP violation. The LHC will also devote a number of runs to accelerate heavy ions and the ALICE collaboration will study their collisions for an experimental exploration of the QCD phase diagram. Other experiments like TOTEM and LHCf complete the exciting LHC programme. By now new input from experiment is badly needed for real progress in particle physics!

\section{QCD}

QCD stands as a main building block of the SM of particle physics. There are no essential problems of principle in its foundations and the comparison with experiment is excellent. 
For many years the relativistic quantum field theory of reference was QED, but at present QCD offers a more complex and intriguing theoretical laboratory. Indeed, due to asymptotic freedom, QCD can be considered as a better defined theory than QED. The statement that QCD is an unbroken renormalizable gauge theory, based on the $SU(3)$ colour group, with six kinds of triplets quarks with given masses, completely specifies the form of the Lagrangian in terms of quark and gluon fields. From the compact form of its Lagrangian one might be led to  think that QCD is a "simple" theory. But actually this simple theory has an extremely rich dynamical content, including the striking properties of asymptotic freedom and of confinement,  the complexity of the observed hadronic spectrum (with light and heavy quarks), the spontaneous breaking of (approximate) chiral symmetry, a complicated phase transition structure (deconfinement, chiral symmetry restoration, colour superconductivity), a highly non trivial vacuum topology (instantons, $U(1)_A$ symmetry breaking, strong CP violation,....), and so on.

So QCD is a complex theory and it is difficult to make its content explicit. Different routes have been developed over the years. There are non perturbative methods: lattice simulations (in great continuous progress), effective lagrangians valid in restricted specified domains, like chiral lagrangians, heavy quark effective theories, Soft Collinear Effective Theories (SCET), Non Relativistic QCD....) and also QCD sum rules, potential models (for quarkonium) etc. But the perturbative approach, based on asymptotic freedom and only applicable to hard processes, still remains the main quantitative connection to experiment. Great experimental work on testing QCD has been accomplished over the years at HERA, LEP and the Tevatron and elsewhere. All of this is very important for the LHC preparation: understanding QCD processes is an essential prerequisite for all possible discoveries. 

Due to confinement no free coloured particles are observed but only colour singlet hadrons. In high energy collisions the produced quarks and gluons materialize as narrow jets of hadrons. Our understanding of the confinement mechanism has much improved thanks to lattice simulations of QCD at finite temperatures and densities \cite{OPhil}.  The potential between two colour charges, obtained from the lattice computations, clearly shows a linear slope at large distances (linearly rising potential). The slope decreases with increasing temperature until it vanishes at a critical temperature $T_C$. Above $T_C$ the slope remains zero. The phase transitions of colour deconfinement and of chiral restoration appear to happen together on the lattice. Near the critical temperature for both deconfinement and chiral restoration a rapid transition is observed in lattice simulations. In particular the energy density $\epsilon(T)$ is seen to sharply increase. The critical parameters and the nature of the phase transition depend on the number of quark flavours $N_f$ and on their masses. For example, for  $N_f$ = 2 or 2+1 (i.e. 2 light u and d quarks and 1 heavier s quark), $T_C \sim 175~MeV$  and $\epsilon(T_C) \sim 0.5-1.0 ~GeV/fm^3$. For realistic values of the masses $m_s$ and $m_{u,d}$ the phase transition appears to be a second order one, while it becomes first order for very small or very large $m_{u,d,s}$. At high densities the colour superconducting phase is probably also present with diquarks acting as Cooper pairs. The hadronic phase and the deconfined phase are separated by a crossover line at small densities and by a critical line at high densities. Determining the exact location of the critical point in T and $\mu_B$ is an important challenge for theory and is also important for the interpretation of heavy ion collision experiments.  

A large investment is being done in experiments of heavy ion collisions \cite{BCole} with the aim of finding some evidence of the quark gluon plasma phase. Many exciting results have been found at the CERN SPS in the past  years and more recently at RHIC \cite{RVenu}. At the CERN SPS some experimental hints of rapid variation of measured quantities with the energy density were found in the form, for example, of $J/ \Psi$ production suppression or of strangeness enhancement when going from p-A to Pb-Pb collisions. Indeed a posteriori the CERN SPS appears well positioned in energy to probe the transition region, in that a marked variation of different observables was observed. One impressive effect detected at RHIC, interpreted as due to the formation of a hot and dense bubble of matter, is the observation of a strong suppression of back-to-back correlations in jets from central collisions in Au-Au, showing that the jet that crosses the bulk of the dense region is absorbed. The produced hot matter shows a high degree of collectivity, as shown by the observation of elliptic flow (produced hadrons show an elliptic distribution while it would be spherical for a gas) and resembles a perfect liquid with small or no viscosity. Elliptic flow, inclusive spectra, partonic energy loss in medium, strangeness enhancement, J/$\Psi$  suppression etc. are all suggestive (but only suggestive!) of early production of a coloured partonic medium with high energy density
and temperature, close to the theoretically expected values, then expanding as a near ideal fluid. The experimental programme on heavy ion collisions will continue at the LHC where ALICE, the dedicated heavy ion collision experiment, is ready to take data with heavy ion beams.

As we have seen, a main approach to non perturbative problems in QCD is by simulations of the theory on the lattice \cite{AKura}, a technique started by K. Wilson in 1974 which has shown continuous progress over the last decades by going to smaller lattice spacing and larger lattices. A recent big step, made possible by the availability of more powerful dedicated computers, is the evolution from quenched (i.e. with no dynamical fermions) to unquenched calculations. Calculations with dynamical fermions (which take into account the effects of virtual quark loops) imply the evaluation of the quark determinant which is a difficult task. How difficult depends on the particular calculation method. There are several approaches (Wilson, twisted mass,  Kogut-Susskind staggered, Ginsparg-Wilson fermions), each with its own advantages and disadvantages (including the time it takes to run the simulation on a computer). Another important progress is in the capability of doing the simulations with lighter quark masses (closer to the physical mass).  As lattice simulations are always limited to masses of light quarks larger than a given value, going to lighter quark masses makes the use of chiral extrapolations less important (to extrapolate the results down to the physical pion mass one can take advantage of the chiral effective theory in order to control the chiral logs: $\log(m_q/4\pi f_\pi)$).  With the progress from unquenching and lighter quark masses an evident improvement in the agreement of predictions with the data is obtained. For example, modern simulations reproduce the hadron spectrum quite well.  For lattice QCD one is now in an epoch of pre-dictivity as opposed to the  post-dictivity of the past. And in fact the range of precise lattice results currently includes many domains including the quark masses, the form factors for K and D decay, the B parameter for kaons, the decay constants $f_K$, $f_D$, $f_{Ds}$, the $B_c$ mass and many more.

We  now discuss  perturbative QCD \cite{GSal}. In the QCD Lagrangian quark masses are the only parameters with dimensions. Naively (or classically) one would expect massless QCD to be scale invariant so that dimensionless observables would not depend on the absolute energy scale but only on ratios of energy variables. While massless QCD in the quantum version, after regularisation and renormalisation, is finally not scale invariant, the theory is asymptotically free and all the departures from scaling are asymptotically small and computable in terms of the running coupling $\alpha_s(Q^2)$ that decreases logarithmically at large $Q^2$. Mass corrections and non perturbative effects, present in the realistic case, are suppressed by powers of $1/Q^2$. 

The measurements of $\alpha_s(Q^2)$  are among the main quantitative tests of the theory. The most  precise and  reliable determinations are from $e^+e^-$ colliders (mainly at LEP: inclusive Z decays, inclusive
hadronic $\tau$ decay, event shapes and jet rates) and from scaling violations in Deep Inelastic Scattering (DIS).  There is a remarkable agreement among these different determinations. An all-inclusive average $\alpha_s(m_Z^2)=0.1184(7)$ is obtained in \cite{beth}, a value which corresponds to $\Lambda_{QCD}\sim 213(9)~MeV$ ($\bar{MS}$, 5 flavours).

Since $\alpha_s$ is not too small, $\alpha_s(m_Z^2) \sim 0.12$, the need of high
order perturbative calculations, of resummation of logs at all 
orders etc. is particularly acute.  Ingenious new computational techniques and software have been developed and many calculations have been realized that only a decade ago appeared as impossible \cite{GSal}.  An increasing number of processes of interest for the physics at the LHC have been computed at NLO.  Methods for the automated calculation of NLO processes have been very much advanced, based on generalised unitarity \cite{uni} and algebraic reduction to basic integrals at the integrand level \cite{opp}.  Powerful tools have been developed for automatic NLO calculations like HELAC, CutTools, BlackHat, Rocket \cite{pit}. Recent examples are the NLO calculations for $p \bar p \rightarrow t \bar t b \bar b$ \cite{pozzo}, for $W + $ 3 jets \cite{w3j} and for $W + $ 4 jets \cite{w4j}.

NNLO calculations are needed for benchmark measurements where experimental errors are small and corrections are large. A number of these extremely sophisticated calculations have been completed. 
In 2004 the complete calculation of the NNLO splitting functions has been published \cite{ref:moc} $\alpha_s P \sim \alpha_s P_1+ \alpha_s^2 P_2 + \alpha_s^3 P_3+\dots$, a really monumental, fully analytic, computation. More recently the main part of the inclusive hadronic $Z$ and $\tau$ decays at $o(\alpha_s^4)$ (NNNLO!) has been computed \cite{ref:bck}. Another (for the LHC) very important example is Higgs production via $g ~+~ g \rightarrow H$ \cite{bol}. The amplitude is dominated by the top quark loop (if heavier coloured particles exist, like quarks of a 4th generation, for example, they would also contribute). The NLO corrections turn out to be particularly large. Higher order corrections can be computed either in the effective lagrangian approach, where the heavy top is integrated away and the loop is shrunk down to a point (the coefficient of the effective vertex is known to $\alpha_s^4$ accuracy), or in the full theory. At the NLO the two approaches agree very well for the rate as a function of $m_H$. The NNLO corrections have been computed in the effective vertex approximation. Beyond fixed order, resummation of large logs were carried out. Also the NLO EW contributions are known by now. Rapidity (at  NNLO) and $p_T$ distributions (at NLO) have also been evaluated. At smaller $p_T$ the large logarithms $[log(p_T/m_H)]^n$ have been resummed in analogy with what was done long ago for W and Z production. 

The precise knowledge of parton density functions (PDF) \cite{heralhc} is instrumental for computing cross-sections of hard processes at hadron colliders via the factorisation formula. The predictions for cross sections and distributions at $pp$ or $p\bar p$ colliders for large $p_T$ jets or photons, for heavy quark production, for Drell-Yan, W and Z production are all in very good agreement with experiment. There was an apparent problem for b quark production at the Tevatron, but the problem appears now to be solved by a combination of refinements (log resummation, B hadrons instead of b quarks, better fragmentation functions....)\cite{ref:cac}. The QCD predictions are so solid that W and Z production are actually considered as possible luminosity monitors for the LHC. 

The activity on event simulation also received a big boost from the LHC preparation (see, for example, \cite{am} and the review \cite{ref:LHC}). General algorithms for performing NLO calculations numerically (requiring techniques for the cancellation of singularities between real and virtual diagrams) have been developed (see, for example, \cite{ref:num}). The matching of matrix element calculation of rates together with the modeling of parton showers has been realised in packages, as for example in the MC@NLO \cite{ref:frix} or POWHEG \cite{ref:fnr} based on HERWIG. The matrix element calculation, improved by resummation of large logs, provides the hard skeleton (with large $p_T$ branchings) while the parton shower is constructed by a sequence of factorized collinear emissions fixed by the QCD splitting functions. In addition, at low scales a model of hadronisation completes the simulation. The importance of all the components - matrix element, parton shower and hadronisation - can be appreciated in simulations of hard events compared with the Tevatron data. 

Important work on jet recombination algorithms has been published by G. Salam and collaborators (for a review, see \cite{GSal}). In fact it is essential that a correct jet finding is implemented by LHC experiments for an optimal matching of theory and experiment. A critical reappraisal of the existing cone and recombination methods has led to new improved versions of jet defining algorithms, like SISCone and anti-$k_T$ with good infra red properties and leading to a simpler jet structure.

In conclusion, I think that the domain of QCD appears as one of great maturity but also of robust vitality (as apparent by the large amount of work produced for the LHC preparation) and all the QCD predictions that one was able to formulate and to test are in good agreement with experiment.

\section{Electroweak interactions and the Higgs problem}

The experimental verification of the SM cannot be considered complete until the predicted physics of the  Higgs sector is not established by experiment \cite{djou1}. Indeed the Higgs problem is really central in particle physics today \cite{wells}. In fact, the Higgs sector is directly related to most of the major open problems of particle physics, like the flavour problem and the hierarchy problem, the latter strongly suggesting the need for new physics near the weak scale, which could also clarify the Dark Matter identity. 

It is clear that the fact that some sort of Higgs mechanism is at work has already been established. The W and Z longitudinal degrees of freedom are borrowed from the Higgs sector and are an evidence for it. In fact the couplings of quarks and leptons to
the weak gauge bosons W$^{\pm}$ and Z are indeed experimentally found to be precisely those
prescribed by the gauge symmetry.  To a lesser
accuracy the triple gauge vertices $\gamma$WW and ZWW have also
been found in agreement with the specific predictions of the
$SU(2)\bigotimes U(1)$ gauge theory. This means that it has been
verified that the gauge symmetry is unbroken in the vertices of the
theory: all currents and charges are indeed symmetric. Yet there is obvious
evidence that the symmetry is instead badly broken in the
masses. The W or the Z with longitudinal polarization that are observed are not present in an unbroken gauge theory (massless spin-1 particles, like the photon, are transversely polarized). Not only the W and the Z have large masses, but the large splitting of, for example,  the t-b doublet shows that even the global weak SU(2) is not at all respected by the fermion spectrum. Symmetric couplings and totally non symmetric spectrum is a clear signal of spontaneous
symmetry breaking and its implementation in a gauge theory is via the Higgs mechanism. The big remaining questions are about the nature and the properties of the Higgs particle(s). 

The LHC has been designed to solve the Higgs problem. A strong argument indicating that the solution of the Higgs problem cannot be too far away is the fact that, in the absence of a Higgs particle or of an alternative mechanism, violations of unitarity appear in scattering amplitudes involving longitudinal gauge bosons (those most directly related to the Higgs sector) at energies in the few TeV range \cite{unit}. A crucial question for the LHC is to identify the mechanism that avoids the unitarity violation: is it one or more Higgs bosons or some new vector boson (like additional gauge bosons $W^\prime$, $Z^\prime$ or Kaluza-Klein recurrences or resonances from a strong sector)?

It is well known that in the SM with only one Higgs doublet a lower limit on
$m_H$ can be derived from the requirement of vacuum stability (i.e. that the quartic Higgs coupling $\lambda$ does not turn negative in its running up to a large scale $\Lambda$) or, in milder form, of a moderate instability, compatible with the lifetime of the Universe  \cite{isi}. The Higgs mass enters because it fixes the initial value of the quartic Higgs coupling $\lambda$. Given the experimental value of $m_t$, for $\Lambda \sim $ a few TeV, the lower limit on $m_H$ is below the direct experimental bound while for $\Lambda \sim M_{Pl}$ it is given by $m_H > 130$ GeV. Similarly an upper bound on $m_H$ (with mild dependence
on $m_t$) is obtained, as described in \cite{hri}, from the requirement that no Landau pole appears, up to the scale $\Lambda$, in the Higgs quartic coupling $\lambda$, or in simpler terms, that the perturbative description of the theory remains valid up to  $\Lambda$. Even if $\Lambda$ is as small as ~a few TeV the limit is well within the LHC range $m_H < 600-800~$GeV and becomes $m_H < 180~$GeV for $\Lambda \sim M_{Pl}$. The upper limit on the Higgs mass in the SM makes it clear that the LHC will say yes or no to the simplest Higgs sector of the SM. 

In conclusion it looks very likely that the LHC can very much advance our knowledge of the electroweak symmetry breaking mechanism. It has been designed for it!

\section{Precision Tests of the Standard Electroweak Theory}

The most precise tests of the electroweak theory apply to the QED sector. The anomalous magnetic moments of the electron and of the muon are among the most precise measurements in the whole of physics. Recently there have been new precise measurements of $a$ for the electron \cite{ae1} and the muon \cite{amu} ($a = (g-2)/2$). The QED part has been computed analytically for $i=1,2,3$, while for $i=4$ there is a numerical calculation (see, for example, \cite{kino}). Some terms for $i=5$ have also been estimated for the muon case. The weak contribution arises from $W$ or $Z$ exchange. The hadronic contribution is from vacuum polarization insertions and from light by light scattering diagrams.  For the electron case the weak contribution is essentially negligible and the hadronic term does not introduce an important uncertainty.  As a result the $a_e$ measurement can be used to obtain the most precise determination of the fine structure constant \cite{ae2}. In the muon case the experimental precision is smaller by about 3 orders of magnitude, but the sensitivity to new physics effects is typically increased by a factor $(m_\mu/m_e)^2 \sim 4^.10^4$. The dominant theoretical ambiguities arise from the hadronic terms in vacuum polarization and in light by light scattering. If the vacuum polarization terms are evaluated from the $e^+e^-$ data a discrepancy of $\sim 3 \sigma$ is obtained  (the $\tau$ data would indicate better agreement, but the connection to $a_\mu$ is less direct and recent new data have added solidity to the $e^+e^-$ route) \cite{amu2}. Finally, we note that, given the great accuracy of the $a_\mu$ measurement and the estimated size of the new physics contributions, for example from SUSY, it is not unreasonable that a first signal of new physics would appear in this quantity.

The results of the electroweak precision tests also including the measurements of $m_t$, $m_W$ and the searches for new physics at the Tevatron form a very stringent set of precise constraints \cite{ewg} to compare with the SM or with
any of its conceivable extensions. When confronted with these results, on the whole the SM performs rather
well, so that it is fair to say that no clear indication for new physics emerges from these data \cite{AG}.  But the
Higgs sector of the SM is still very much untested. What has been
tested is the relation $M_W^2=M_Z^2\cos^2{\theta_W}$, modified by small, computable
radiative corrections. This relation means that the effective Higgs
(be it fundamental or composite) is indeed a weak isospin doublet.
The Higgs particle has not been found but in the SM its mass can easily
be larger than the present direct lower limit $m_H > 114.4$~GeV
obtained from direct searches at LEP-2 \cite{ewg}.  The radiative corrections
computed in the SM when compared to the data on precision electroweak
tests lead to a clear indication for a light Higgs, not too far from
the present lower bound. The exact upper limit for $m_H$ in the SM depends on the value of the top quark mass $m_t$ (the one-loop radiative corrections are quadratic in $m_t$ and logarithmic in $m_H$). The CDF and D0 combined value is at present $m_t~= 173.3~\pm~1.1~GeV$. As a consequence the present limit on $m_H$ is quite stringent: $m_H < 185~GeV$ (at $95\%$ c.l., after including the information from the 114.4 GeV direct bound)  \cite{ewg}.  

In the Higgs search the Tevatron has now reached the SM sensitivity. The most recent limit, reported at ICHEP '10, is: $158 < m_H < 175$ GeV \cite{BKilminster}. This quoted $95\%$ c.l. limit of course depends on the assumptions made on the Higgs production cross sections and branching ratios.

\section{The Physics of Flavour}

Another domain where the SM is in very good agreement with the data is flavour physics (really too good in comparison with the general expectation before the experiments). In the last decade great progress in different areas of flavour physics has been achieved. In the quark sector, the amazing results of a generation of frontier experiments, performed at B factories and at accelerators, have become available. 
The hope of the B-decay experiments was to detect departures from the CKM picture of mixing and of CP violation as  signals of new physics. At present the available results on B mixing and CP violation on the whole agree very well with the SM predictions based on the CKM matrix \cite{rept}. A few interesting ÒtensionsÓ at the 2-3 $\sigma$ level should be monitored closely in the future (in particular by LHCb): $\sin{2\beta}$ from $B_d \rightarrow J/\Psi K^0$ versus $\epsilon_K$ and $V_{ub}$ (which, however, in my opinion, is probably due to an underestimate of theoretical errors, particularly on the determination of $V_{ub}$), $\beta_s$ measured by CDF and $D_0$
in $B_s \rightarrow J/\Psi \phi$, the D0 dimuon asymmetry interpreted as arising from CP violation in $B_s$ semileptonic decay and $B\rightarrow \tau \nu$. But certainly the amazing performance of the SM in flavour changing neutral current  and/or CP violating transitions in K and B decays poses very strong constraints on all proposed models of new physics \cite{isid}. For example, if to the SM one adds effective non renormalizable operators suppressed by powers of a scale $\Lambda$, with coefficients of $o(1)$, one generically finds that experiments point to very large values of $\Lambda$, much above the few TeV range indicated by the hierarchy problem.  In order to obtain bounds on  $\Lambda$ in the few TeV range one has to assume that the relevant new physics effects are suppressed at the tree level and mainly occur at the loop level and that, in addition, the new physics inherits the same SM protections against flavour changing neutral currents, like the GIM mechanism and small $V_{CKM}$ factors, as, for example, in Minimal Flavour Violation models \cite{isid}.

In the leptonic sector the study of neutrino oscillations has led to the discovery that at least two neutrinos are not massless and to the determination of the mixing matrix \cite{revnu}. Neutrinos are not all massless but their masses are very small (at most a fraction of $eV$). The neutrino spectrum could be either of the normal hierarchy type (with the solar doublet below), or of the inverse hierarchy type (with the solar doublet above). Probably masses are small because $\nu^\prime$s are Majorana fermions, and, by the see-saw mechanism, their masses are inversely proportional to the large scale $M$ where lepton number ($L$) non conservation occurs (as expected in GUT's). Indeed the value of $M\sim m_{\nu R}$ from experiment is compatible with being close to $M_{GUT} \sim 10^{14}-10^{15}GeV$, so that neutrino masses fit well in the GUT picture and actually support it. The interpretation of neutrinos as Majorana particles enhances the importance of experiments aimed at the detection of neutrinoless double beta decay and a huge effort in this direction is underway.  It was realized that decays of heavy $\nu_R$ with CP and L non conservation can produce a B-L asymmetry (which is unchanged by instanton effects at the electroweak scale). The range of neutrino masses indicated by neutrino phenomenology turns out to be perfectly compatible with the idea of baryogenesis via leptogenesis \cite{buch}. This elegant model for baryogenesis has by now replaced the idea of baryogenesis near the weak scale, which has been strongly disfavoured by LEP. It is remarkable that we now know the neutrino mixing matrix with good accuracy \cite{datanu}. Two mixing angles are large and one is small. The atmospheric angle $\theta_{23}$ is large, actually compatible with maximal but not necessarily so. The solar angle $\theta_{12}$ (the best measured) is large, $\sin^2{\theta_{12}}\sim 0.3$, but certainly not maximal. The third angle $\theta_{13}$, strongly limited mainly by the CHOOZ experiment, has at present a $3\sigma$ upper limit given by about $\sin^2{\theta_{13}}\leq 0.04$. It is a fact that, to a precision comparable with the measurement accuracy, the Tri-Bimaximal (TB) mixing pattern ($\sin^2{\theta_{12}}\sim 1/3$, $\sin^2{\theta_{23}}\sim 1/2$ and $\sin^2{\theta_{13}} = 0$) \cite{Harr} is well approximated by the data. If this experimental result is not a mere accident but a real indication that a dynamical mechanism is at work to guarantee the validity of TB mixing in the leading approximation, corrected by small non leading terms, then non abelian discrete flavor groups emerge as the main road to an understanding of this mixing pattern \cite{rmp}. Indeed the entries of the TB mixing matrix are suggestive of "rotations" by simple, very specific angles. In fact the group $A_4$, the simplest group used to explain TB mixing, is defined as the group of rotations that leave a regular rigid tetrahedron invariant. The non conservation of the three separate lepton numbers and the large leptonic mixing angles make it possible that processes like $\mu \rightarrow e \gamma$ or $\tau \rightarrow \mu \gamma$ which are beyond reach in the SM could be observable in extensions of it like the MSSM. Thus, the outcome of the now running experiment MEG at PSI \cite{MEG} aiming at improving the limit on $\mu \rightarrow e \gamma$ by 1 or 2 orders of magnitude, is of particular interest. 

\section{Outlook on Avenues beyond the Standard Model}
\label{sec:5}

No signal of new physics has been
found neither in electroweak precision tests nor in flavour physics. Given the success of the SM why are we not satisfied with that theory? Why not just find the Higgs particle,
for completeness, and declare that particle physics is closed? The reason is that there are
both conceptual problems and phenomenological indications for physics beyond the SM. On the conceptual side the most
obvious problems are the proliferation of parameters, the puzzles of family replication and of flavour hierarchies, the fact that quantum gravity is not included in the SM and the related hierarchy problem. Among the main
phenomenological hints for new physics we can list the constraints from coupling constant merging in Grand Unified Theories (GUT's), dark matter, neutrino masses (explained in terms of L non conservation), 
baryogenesis and the cosmological vacuum energy (a gigantic naturalness problem).

The computable evolution with energy
of the effective gauge couplings clearly points towards the unification of the electro-weak and strong forces at scales of energy
$M_{GUT}\sim  10^{15}-10^{16}~ GeV$ which are close to the scale of quantum gravity, $M_{Pl}\sim 10^{19}~ GeV$.  One is led to
imagine  a unified theory of all interactions also including gravity (at present superstrings provide the best attempt at such
a theory). Thus GUT's and the realm of quantum gravity set a very distant energy horizon that modern particle theory cannot
ignore. Can the SM without new physics be valid up to such large energies? Some of the SM problems could be postponed to the more fundamental theory at the Planck mass. For example, the explanation of the three generations of fermions and the understanding of fermion masses and mixing angles can be postponed. But other problems must find their solution in the low energy theory. In particular, the structure of the
SM could not naturally explain the relative smallness of the weak scale of mass, set by the Higgs mechanism at $\mu\sim
1/\sqrt{G_F}\sim  250~ GeV$  with $G_F$ being the Fermi coupling constant. This so-called hierarchy problem is due to the instability of the SM with respect to quantum corrections. This is related to
the
presence of fundamental scalar fields in the theory with quadratic mass divergences and no protective extra symmetry at
$\mu=0$. For fermion masses, first, the divergences are logarithmic and, second, they are forbidden by the $SU(2)\bigotimes
U(1)$ gauge symmetry plus the fact that at
$m=0$ an additional symmetry, i.e. chiral  symmetry, is restored. Here, when talking of divergences, we are not
worried of actual infinities. The theory is renormalizable and finite once the dependence on the cut-off $\Lambda$ is
absorbed in a redefinition of masses and couplings. Rather the hierarchy problem is one of naturalness. We can look at the
cut-off as a parameterization of our ignorance on the new physics that will modify the theory at large energy
scales. Then it is relevant to look at the dependence of physical quantities on the cut-off and to demand that no
unexplained enormously accurate cancellations arise. 

The hierarchy problem can be put in less abstract terms: loop corrections to the Higgs mass squared are
quadratic in the cut-off $\Lambda$. The most pressing problem is from the top loop (the heaviest particle, hence the most coupled to the Higgs).
 With $m_h^2=m^2_{bare}+\delta m_h^2$ the top loop gives 
 \begin{eqnarray}
\delta m_{h|top}^2\sim -\frac{3G_F}{2\sqrt{2} \pi^2} m_t^2 \Lambda^2\sim -(0.2\Lambda)^2 \label{top}
\end{eqnarray}
If we demand that the correction does not exceed the light Higgs mass indicated by the precision tests, $\Lambda$ must be
close, $\Lambda\sim o(1~TeV)$. So a crucial question for the LHC to answer is: what damps the top loop contribution? Similar constraints arise from the quadratic $\Lambda$ dependence of loops with gauge bosons and
scalars, which, however, lead to less pressing bounds. So the hierarchy problem demands new physics to be very close (in
particular the mechanism that quenches the top loop). Actually, this new physics must be rather special, because it must be
very close, yet its effects are not clearly visible (the "LEP Paradox" \cite{BS}) now also accompanied by a similar "flavour paradox" \cite{isid}. Examples of proposed classes of solutions
for the hierarchy problem are listed in the following \cite{CGrojean}.

¥ $\bf{Supersymmetry (SUSY).}$ In the limit of exact boson-fermion symmetry \cite{Martin} the quadratic divergences of bosons cancel so that
only log divergences remain. However, exact SUSY is clearly unrealistic. For approximate SUSY (with soft breaking terms),
which is the basis for all practical models, $\Lambda$ is replaced by the splitting of SUSY multiplets, $\Lambda\sim
m_{SUSY}-m_{ord}$. In particular, the top loop is quenched by partial cancellation with s-top exchange, so the s-top cannot be too heavy. Important phenomenological indications in favour of SUSY are that coupling unification takes place with greater accuracy in SUSY than in the SM and that proton decay bounds are not in contradiction with the predictions. Two Higgs doublets are expected in SUSY and the lightest Higgs should be really light: $m_h < 130$ GeV \cite{djou2}. An interesting exercise is to repeat the fit of precision tests in the Minimal Supersymmetric Standard Model (MSSM) with GUT constraints added, also including the additional data on the muon $(g-2)$, the dark matter relic density and on the $b\rightarrow s \gamma$ rate. The result is that the central value of the lightest Higgs mass $m_h$ goes up (in better harmony with the bound from direct searches) with moderately large $tan\beta$ and relatively light SUSY spectrum \cite{OBuch}. The problem with SUSY is that one expected its discovery already at LEP2 on the basis of complete naturalness applied to minimal models. However less fine tuning is necessary if non minimal models are assumed (for a recent example, see \cite{barbie}).  

¥ $\bf{Technicolor.}$ In these models the Higgs system is a condensate of new fermions. There is no fundamental scalar Higgs sector, hence no
quadratic divergences associated to the $\mu^2$ mass in the scalar potential. But this mechanism needs a very strong binding force,
$\Lambda_{TC}\sim 10^3~\Lambda_{QCD}$ and, as a consequence, it is  difficult to arrange that such nearby strong force is not showing up in
precision tests. Hence this class of models has been disfavoured by LEP, although some special versions that can possibly work have been a-posteriori formulated, like walking TC, top-color assisted TC etc \cite{L-C} and more recently some extra dimensional models based on the AdS/CFT correspondence \cite{con}. 

¥ $\bf{"Little~Higgs"~models.}$  In "little Higgs" models \cite{schm} the symmetry of the SM is extended to a suitable global group G that also contains some
gauge enlargement of $SU(2)\bigotimes U(1)$, for example $G\supset [SU(2)\bigotimes U(1)]^2\supset SU(2)\bigotimes
U(1)$. The Higgs particle is a pseudo-Goldstone boson of G that can only take mass at the 2-loop level, because two distinct
symmetries must be simultaneously broken for this to happen, which requires the action of two different couplings in
the same diagram. Then in the relation eq.(\ref{top})
between
$\delta m_h^2$ and
$\Lambda^2$ there are an additional coupling and an additional loop factor that imply a larger separation between the Higgs
mass and the cut-off. Typically, in these models one has one or more Higgs doublets at $m_h\sim~0.2~{\rm TeV}$, and a cut-off at
$\Lambda\sim~10~{\rm TeV}$. The top loop quadratic cut-off dependence is partially canceled, in a natural way guaranteed by the
symmetries of the model, by a new coloured, charge 2/3, vectorlike quark $\chi$ of mass around $1~{\rm TeV}$ (a fermion not a scalar
like the s-top of SUSY models). Certainly these models involve a remarkable level of group theoretic virtuosity. However, in
the simplest versions one is faced with problems with precision tests of the SM \cite{prob}. These problems can be fixed by complicating the model \cite{Ch}: one can introduce a parity symmetry, T-parity, and additional "mirror" fermions.  T-parity interchanges the two $SU(2)\bigotimes
U(1)$ groups: standard gauge bosons are T-even while heavy ones are T-odd. As a consequence no tree level contributions from heavy $WÕ$ and $ZÕ$ appear in processes with external SM particles. 
Therefore all corrections to EW observables only arise at loop level. A good feature of T-parity is that, like for R-parity in the MSSM, the lightest T-odd particle is stable (usually a B') and can be a candidate for Dark Matter (missing energy would here too be a signal) and T-odd particles are produced in pairs (unless T-parity is not broken by anomalies \cite{hill}). Thus the model could work but, in my opinion, the real limit of
this approach is that it only offers a postponement of the main problem by a few TeV, paid by a complete loss of
predictivity at higher energies. In particular all connections to GUT's are lost. Still it is very useful as it offers to experiment a different example of possible new physics and the related signals to look for \cite{sign}.

¥ $\bf{Extra~dimensions.}$ In the original approach \cite{led} the idea was that $M_{Pl}$ appears very large, or equivalently that gravity appears very weak,
because we are fooled by hidden extra dimensions (ED), so that the real gravity scale is reduced down to a much lower scale and that effects of extra dimensions could be detectable at energies of
$o(1~TeV)$ ("large" extra dimensions). This possibility is very exciting in itself and it is really remarkable that it is not directly incompatible with experiment but a realistic model has not emerged \cite{Jo}. At present, the most promising set of ED models are those with "warped" metric, which offer attractive solutions to the hierarchy problem \cite{RS}, \cite{FeAa}. The hierarchy suppression $m_W/M_{Pl}$ arises from the warping exponential $e^{-kR\phi}$, with $k\sim M_{Pl}$, for not too large values of the warp factor exponent: $kR\sim 12$ (ED are not "large" in this case). The question of whether these values of $kR$ can be stabilized has been discussed in ref.\cite{GW}. An important direction of development is the study of symmetry breaking by orbifolding and/or boundary conditions. These are models where a larger gauge symmetry (with or without SUSY) holds in the bulk. The symmetry is reduced on the 4-dim. brane, where the physics that we observe is located, as an effect of symmetry breaking induced geometrically by orbifolding or by suitable boundary conditions. In particular SUSY GUT models in ED have been studied where the breaking of the GUT symmetry by orbifolding avoids the introduction of large Higgs representations and also solves the doublet-triplet splitting problem \cite{Kaw}, \cite{edgut}. Also "Higgsless  models" have been tried where it is the SM electroweak gauge symmetry which is broken at the boundaries \cite{Hless}. The violation of unitarity associated with the absence of the Higgs exchange is damped by the Kaluza-Klein recurrences of the gauge bosons. In this case no Higgs should be found at the LHC but other signals, like additional vector bosons, should appear.  The main difficulty is represented by the compatibility with the electro-weak precision tests. 

An interesting class of models that combine the idea of the Higgs as a pseudo-Goldstone boson and warped ED was proposed and studied in ref.s \cite{con} where a kind of composite Higgs in a 5-dim AdS theory appears. This approach can be considered as a new way to look at technicolor \cite{L-C} using the AdS/CFT correspondence. In a RS warped metric framework all SM fields are in the bulk but the Higgs is localised near the TeV brane. The Higgs is a pseudo-Goldstone boson and the electroweak symmetry breaking is triggered by 
top-loop effects. In 4-dim the bulk appears as a strong sector.  The 5-dim theory is weakly coupled so that the Higgs potential and EW observables can be computed.
The Higgs is rather light: $m_H < 185~{\rm GeV}$. Problems with EW precision tests and the $Zb \bar b$ vertex have been fixed in latest versions. The signals at the LHC for this model are 
a light Higgs and new resonances at ~1- 2 TeV

In conclusion, note that apart from Higgsless models (if any?) all theories discussed 
here have a Higgs in LHC range (most of them light).

¥ $\bf{Effective~theories~for~compositeness.}$ In this framework \cite {cont}, \cite{comp}, \cite{comp2} a low energy theory from truncation of some UV completion is described in terms of an elementary sector (the SM particles minus the Higgs), a composite sector (including the Higgs, massive vector bosons and new fermions) and a mixing sector. The Higgs is a pseudo-Goldstone boson of a larger broken gauge group. At low energy, the particle content is identical to the SM one: there exists a 
light and narrow Higgs-like scalar but this particle is a composite from some strong dynamics
and a mass gap separates the Higgs boson from the strong sector as 
a result of the Goldstone nature of the Higgs. The 
effective Lagrangian can be seen as an expansion in $\xi = (v/f)^2$  where $v$ is the Higgs vev and $f$ is the typical scale of the strong sector. The parameter $\xi$ interpolates between the SM limit ($\xi = 0$) and the technicolor limit ($\xi = 1$), 
where a resummation of the full series in $\xi$ is needed.
Non vanishing values of $\xi$ can lead to observable signatures of compositeness at the LHC \cite{comp2}). 

¥ $\bf{The~anthropic~solution.}$ The apparent value of the cosmological constant $\Lambda$ poses a tremendous, unsolved naturalness problem \cite{tu}. Yet the value of $\Lambda$ is close to the Weinberg upper bound for galaxy formation \cite{We}. Possibly our Universe is just one of infinitely many (Multiverse) continuously created from the vacuum by quantum fluctuations. Different physics takes place in different Universes according to the multitude of string theory solutions (~$10^{500}$). Perhaps we live in a very unlikely Universe but the only one that allows our existence \cite{anto},\cite{giu}. I find applying the anthropic principle to the SM hierarchy problem excessive. After all we can find plenty of models that easily reduce the fine tuning from $10^{14}$ to $10^2$: why make our Universe so terribly unlikely? By comparison the case of the cosmological constant is a lot different: the context is not as fully specified as the for the SM (quantum gravity, string cosmology, branes in extra dimensions, wormholes through different Universes....)

\section{Conclusion}

Supersymmetry remains the standard way beyond the SM. What is unique to SUSY, beyond leading to a set of consistent and
completely formulated models, as, for example, the MSSM, is that this theory can potentially work up to the GUT energy scale.
In this respect it is the most ambitious model because it describes a computable framework that could be valid all the way
up to the vicinity of the Planck mass. The SUSY models are perfectly compatible with GUT's and are actually quantitatively
supported by coupling unification and also by what we have recently learnt on neutrino masses. Other  ideas for going
beyond the SM do not share this synthesis with GUT's. The SUSY way is testable, for example at the LHC, and the issue
of its validity will be decided by experiment. It is true that we could have expected the first signals of SUSY already at
LEP, based on naturality arguments applied to the most minimal models (for example, those with gaugino universality at
asymptotic scales). The absence of signals has stimulated the development of new ideas like those of extra dimensions
and "little Higgs" models. These ideas are very interesting and provide an important reference for the preparation of LHC
experiments. Models along these new ideas are not so completely formulated and studied as for SUSY and no well defined and
realistic baseline has sofar emerged. But it is well possible that they might represent at least a part of the truth and it
is very important to continue the exploration of new ways beyond the SM. New input from experiment is badly needed, so we all look forward to the start of the LHC.

The future of particle physics heavily depends on the outcome of the LHC. So the questions that many people ask are listed in the following with my (tentative) answers. Is it possible that the LHC does not find the Higgs particle? Yes, it is possible, but then must find something else (experimental and theoretical upper bounds on the Higgs mass in the SM, unitarity violations in the absence of a Higgs or of new physics). Is it possible that the LHC finds the Higgs particle but no other new physics (pure and simple SM)? Yes, it is  technically possible but it is not natural (would go in the direction that we live in a very eccentric Universe). Is it possible that the LHC finds neither the Higgs nor new physics? No, it is Òapproximately impossibleÓ (meaning that the only possible way out would be that the LHC energy is a bit too low and only misses by a small gap the onset of the solution).

\section*{Acknowledgements}
We recognize that this work has been partly supported by the Italian Ministero 
dell'Universit\`a e della Ricerca Scientifica, under the COFIN program (PRIN 2008) 
and by the European Commission
under the network  "Heptools".

%\appendix
%\section{First Appendix} %Empty argument \section{} yields `Appendix'. 
%
%\section{Second Appendix}

\end{document}